\documentclass[twocolumn,showpacs,preprintnumbers,amsmath,amssymb]{revtex4}
\usepackage{epsfig}
\usepackage{graphicx}
\usepackage{dcolumn}
\usepackage{bm}
\begin{document}\hbadness=10000

\title{Hadron resonance probes of QGP}
\author{Giorgio Torrieri and Johann Rafelski}
\affiliation{
Department of Physics, University of Arizona, Tucson, Arizona 85721, USA}
\date{January, 2004}

\begin{abstract}
We discuss the  indirect and direct role of the  short-lived resonances
 as probes of QGP freeze-out process. The indirect effect is the distortion of 
stable single particle yields and spectra by  contributions 
of decaying resonances, which
 alter significantly the parameters obtained in  fits to experimental data.
We than discuss
 the direct observation of short-lived resonances as a
probe of post-hadronization dynamics allowing to 
distinguish between different hadronization models.
\end{abstract}

\vspace{+0.5cm}



\maketitle
\section{Introduction}
We would like to understand the dynamics of
 quark-gluon plasma (QGP) phase breakup
into individual hadrons and show here how this
can be done using hadron resonances. To convert QGP into hadrons we employ 
the Fermi statistical model of particle
 production~\cite{Fer50,Pom51,Lan53,Hag65,CooperFrye,jansbook}.
This approach has been used extensively in the field of
 relativistic heavy ion collisions. 
Particle abundances and spectra both at Super Proton Synchrotron (SPS) 
~\cite{PBM99,NA57,becattini,van-leeuwen,ourspspaper,sqm2001,PBM01} and Relativistic Heavy Ion Collider (RHIC) 
~\cite{burward-hoy,castillo,PBMRHIC,bugaev_freeze,florkowski,rafelski2002} energies have been analyzed in this way. In all this studies one relies on
presence of a distinct condition in temperature 
 at which the rate of particle interactions goes to zero (freeze-out).
The  HBT \cite{fasthbt} measurements and the success of statistical
models suggest that the freeze-out process at SPS and RHIC energies is
indeed much faster than previously expected, 
being close to the explosive hadronization
limit of instantaneous emission and negligible post-emission interactions \cite{sudden}. 

The obvious way to test this model experimentally is to obtain 
experimental measure of the
time lapse between hadronization and end of hadron-hadron interaction.
While HBT has always been considered the ideal probe to do it, the fact that
hydrodynamic models at present fail to describe HBT data \cite{heinzkolb}, together with the problems associated with emission from an interacting gas with non-zero mean free path \cite{wrongHBT} make the search for a different probe of freeze-out time necessary. Since hadronization is a fast process, resonance decay
products have an appreciable chance of escaping without rescattering and 
thus resonance yields can be measured directly, and in an analysis of
these results one can obtain information about hadronization dynamics.

\begin{figure}
\psfig{width=7.8cm,figure=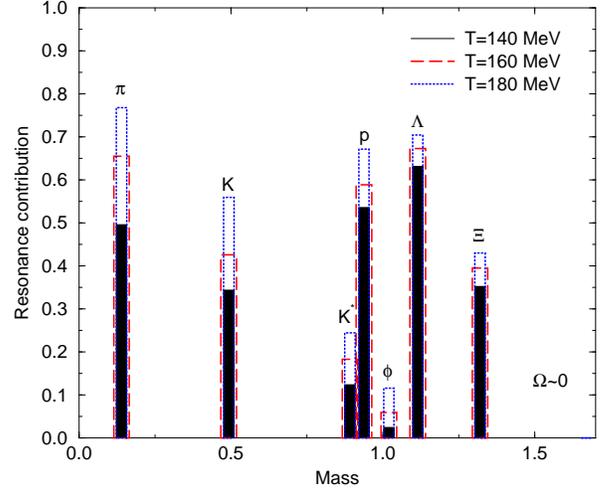}
\caption{(Color online) Relative resonances contribution to 
individual stable hadrons for three particle freeze-out temperatures.
 \label{yieldrs}}
\end{figure}

Therefore we  consider
direct detection of hadronic short-lived resonances as an alternative probe
of freeze-out dynamics. We find that 
the short-lived resonances, detectable through
invariant mass reconstruction~\cite{fachini,vanburen,markert,friese} are natural candidates for freeze-out diagnostics since their lifetime is comparable
to the hadronization timescale and the lifetime of the interacting HG.
Resonances usually have the same quark numbers as light particles, making
their yield compared to the light particle independent of chemical potential.
The rich variety of detected resonances \cite{vanburen} includes particles with very different masses and widths, allowing us to probe both production temperatures
and interaction lifetimes in detail.
Fig. \ref{yieldrs} shows what percentage of observed light particles comes
from the decays of heavier resonances (quite a few of them experimentally observable).
As can be seen in figure \ref{yieldrs} this resonance 
contribution is significant, and varies appreciably with both
particle type and temperature.

\section{Resonance influence  on particle spectra}

Statistical hadronization model assumes that, in the rest-frame with respect to
collective matter flow, particle spectra are given by the
 (entropy maximizing) Boltzmann spectrum.
In addition, the phase space occupancy by particles is  affected 
by the space-time geometry of the particle  emission surface.
The particle  spectrum in the laboratory rest frame is given by,
\begin{eqnarray}
\frac{d^2 N}{dm_{T} dy} &\propto&
\left(1- \frac{ \vec{v}_{f}^{\,-1} \cdot \vec{p}}{E}\right)
 \,  m_{T} \cosh y \,
e^{- \gamma \frac E T \left(1-\frac{\vec{v}\cdot \vec{p}}{E}\right)} \,,\quad \label{cooperfrye}\\
 \gamma&=&1/\sqrt{1-v^2}.
\end{eqnarray}

Taking feed-down from resonances into account can be tedious numerical task.
To simplify the situation we assume that, in a decay of the form,
\begin{equation}
R \rightarrow 1 + 2+...,
\label{dectype}
\end{equation}
dynamical effects in the decay  average out
over a statistical sample of many resonances.
In other words, in the rest frame comoving with the ``average'' 
resonance, the distribution
of the decay products will be isotropic.
If more than two body decays  are considered, this calculation becomes
more involved \cite{phasespace}.
For the general N-body case, evaluation is better left to Monte Carlo methods
\cite{mambo}.

The rate of particles of type 1 (as in eq. (\ref{dectype}))
produced with momentum $\vec{p^{*}_{1}}$ in the frame
at rest w.r.t. the resonance will
then  be given by  the Lorenz invariant phase space factor
of a particle of mass $M_1$ and momentum $\vec{p^{*}_{1}}$ within a system
with center of mass energy equal to the resonance mass $M_R$,
\begin{equation}
\label{decayphase}
\frac{d^3 N_1}{d^3 p^{*}_{1}}=b \int \prod_{i=2}^{n} \frac{d^3 p^*_i}{2 E^*_i} \delta(\sum_{i=2}^N p^*_i-p^*_1) \delta(\sum_{i=2}^N E^*_i-E^*_{1}-M_R).
\end{equation}
Here $b$ is the branching ratio of the considered decay channel.
To obtain laboratory spectrum we need to change coordinates 
from the resonance's  rest frame $(p^*,E^*)$ to
the lab frame $(p,E)$.

In the case of the 2-body decay $p^*,E^*$ are fixed 
by the masses of the decay products,
\begin{eqnarray}
\label{2bodpstar}
E^{*}_1=\frac{1}{2M_R}(M_R^2-m_1^2-m_2^2),\\ 
p^{*}_1=-p^{*}_2=\sqrt{E_1^{*2}-m_1^2}.
\end{eqnarray}
Putting the constraints in eq. (\ref{2bodpstar}) into eq. (\ref{decayphase})
one gets, after some algebra
\cite{resonances}
\begin{eqnarray}
\label{reso}
&&\hspace*{-0.6cm} \frac{dN}{d {m^2_{T1}} d y_1 } =
\frac{ b}{4 \pi p^{*}_1}
\int_{Y-}^{Y+}\!\!\! dY_1\!\!
\int_{M_{T}-}^{M_{T}+}\!\!\! dM_{T1}^{2} J 
\frac{d^2 N_{R}}{dM_{TR}^{2} dY_R} , \\
&&\hspace*{-0.3cm}\noindent J=\frac{M_R}{\sqrt{P_{TR}^2 p_{T1}^2 -(M_R E^{*}_R - M_{TR} m_{T1} 
                                                        \cosh \Delta Y)^2 }},\\
&&\hspace*{-0.6cm}\Delta Y=Y_R-y_1.
\end{eqnarray}
$J$ is  the Jacobian of the transformation from the resonance rest
frame to the lab frame, and the limits of the kinematically allowed integration
region are:
\[\ Y_{\pm}=y_1 \pm \sinh^{-1} \left( \frac{p^{*}_1}{m_{T1}} \right) \]
\[\ M_{T}^{\pm}=M_R 
\frac{E^{*}_R m_{T1} \cosh(\Delta Y) \pm p_{T1} \sqrt{p^{*2}_1-m_{T1}^{2} \sinh^{2} (\Delta Y)}}
{m_{T1}^{2} \sinh^{2} (\Delta Y)+m^{2}_1}\]

\begin{figure}
\epsfig{width=7cm,clip=,figure=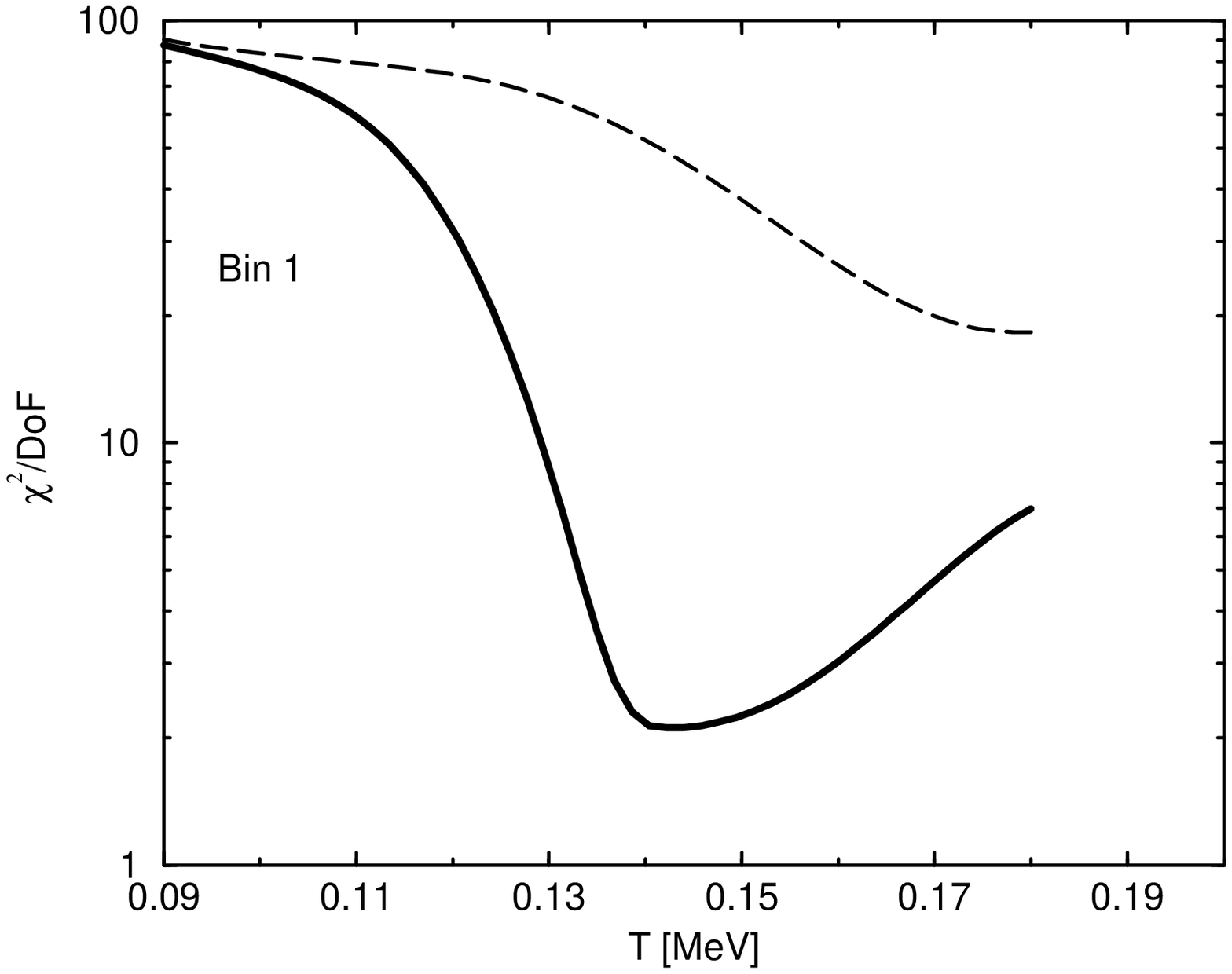}
\epsfig{width=7cm,clip=,figure=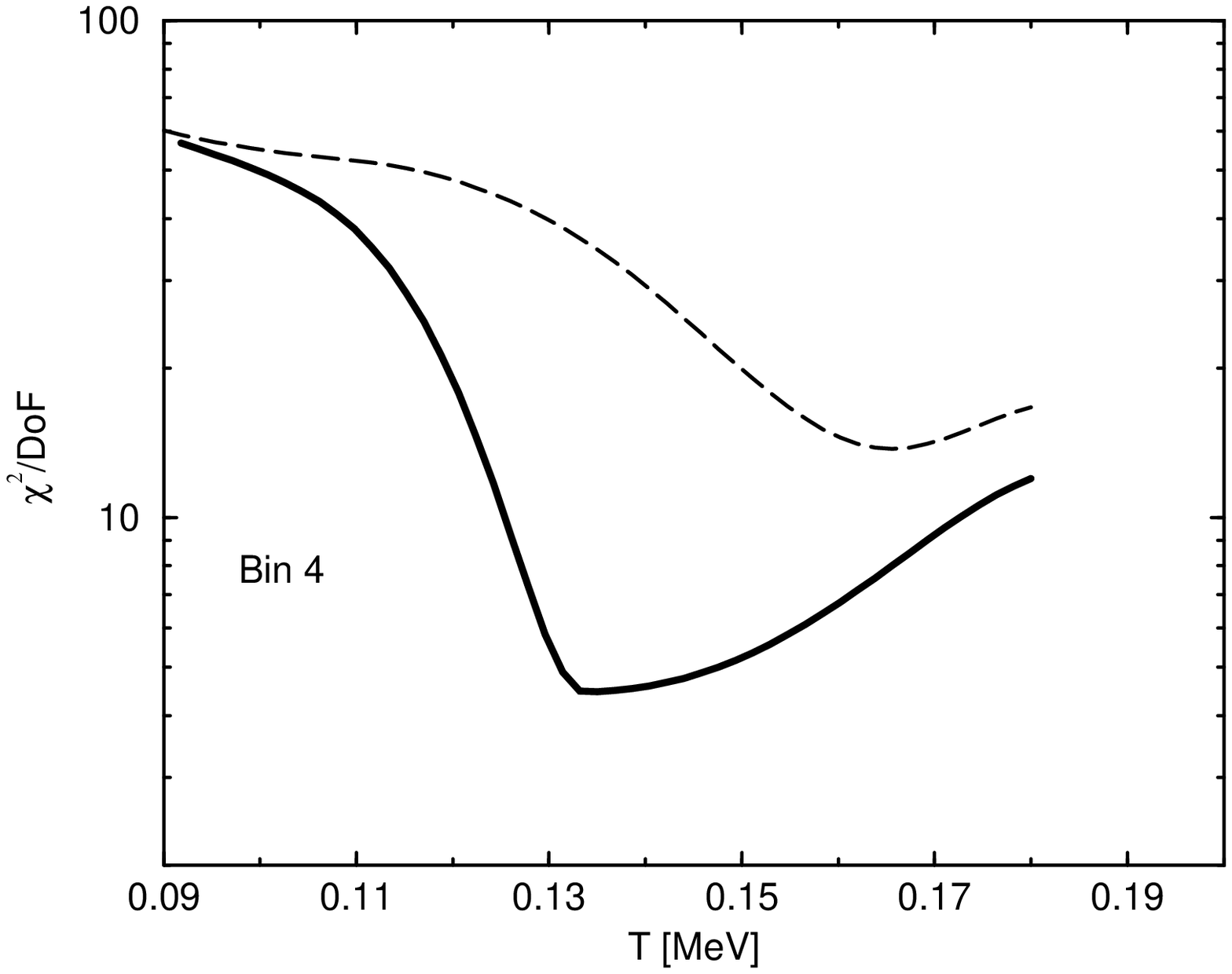}
\caption{ $\chi^2$ profile for a fit to WA97 $K_S,\Lambda,\overline{\Lambda},\Xi,\overline{\Xi},
\Omega,\overline{\Omega}$ data at 158 GeV \cite{wa97}. 
Solid line: resonance decays included.
Dotted line: resonances reequilibrated. \label{chiprofsps}}
\end{figure}

This calculation assumes resonances are produced through the same statistical
model which is fitted to the momenta of the daughter particles, and their
decay products reach the detector with no further interaction. How the data
is described in this approach is seen in figure \ref{chiprofsps} 
for the most peripheral, and the most central reaction bin. Solid lines show
the no-rescattering fit quality for particle spectra.
When rescattering of the decay products is important 
 the shape of the spectra are as if there were 
no resonance decays. This case is presented by dashed lines in  figure \ref{chiprofsps}. Clearly there is a much more physically 
significant spectra fit possible 
if we allow that hadron resonances are present and their decay products
do not rescatter. 

Inclusion of resonances in the spectral fit requires
no extra degrees of freedom, and the improvement of $\chi^2$ of the fit
of the  SPS hyperon spectra seen in figure \ref{chiprofsps} \cite{wa97} 
convincingly favors models in which resonances are present within the
 fireball in the abundances predicted by a single freeze-out temperature 
model.
Not including spectral contributions of resonances raises 
the $\chi^2$ by a factor of 3 and removes  the $\chi^2$ minimum's 
significance. 
We have not yet found a particle type for which consideration 
of  resonances does not lower the $\chi^2$. Moreover, 
we find considering  possible  mass or width shift that such effects 
 significantly increase the $\chi^2$ \cite{ourspspaper}, and thus 
there is another evidence that rescattering, which usually shifts masses and 
widths  \cite{shuryak}, is minimal.

\section{Resonance yields as probe of hadronization dynamics}
The fact that resonances have been found to give the contribution to particle
spectra predicted by the statistical model has motivated the direct search of
resonances through invariant mass reconstruction~\cite{fachini,vanburen,markert,friese}.
It has been found that while statistical models are able to 
fit the $K^*/K$ ratio within error \cite{PBMRHIC}, the $\Lambda(1520)$ 
is strongly suppressed at both SPS and RHIC energies \cite{markert}.
This puzzling result suggests that both emission temperature and
re-scattering might make a significant contribution to resonance abundance, and
their roles need to be disentangles.
This is possible by considering abundances of two resonances with 
different masses (which constrains the temperature) as well as lifetimes (which probes
the role of non-equilibrium in-medium effects, i.e. rescattering).

To obtain a quantitative estimate, we have calculated the ratios of 
$(K^*+\overline{K^*})/K_S$ and
$\Lambda(1520)/\Lambda$ using the statistical model.
In both cases chemical potential corrections are negligible since the 
particle's chemical composition is the same and thus we have:
\begin{eqnarray}
\label{statistical}
\frac{N^*}{N+N^*}&=& \frac{n(m^*,T)}{n(m^*,T)+n(m,T)},\\[0.4cm]
n(m,T)&=& m^2 T K_2 \left( \frac{m}{T} \right).
\end{eqnarray}
We then evolved in time the ratios using a model which combines an average 
rescattering cross-section with dilution due to a constant collective expansion.
In this model, the initial resonances decay with width $\Gamma$ through 
the process $N^* \rightarrow D$.  
Their decay products $D$ ($D(t=0)=0$) then undergo rescattering at a rate
proportional to the medium's density as well as the average rescattering rate.
The final evolution equations then are 
\begin{eqnarray}
\label{rescattering}
 \frac{d N^*}{d t}& =& -\Gamma N^* + R, \\
 \frac{d D}{d t}&=& \Gamma N^* -D\sum_j
\langle \sigma_{Dj} v_{Dj}\rangle \rho_j
          \left(\frac{R_0}{R_0+v t}\right)^3. 
\end{eqnarray}
$v$ is the expansion velocity, $R_0$ is the hadronization radius, 
$\rho_j=  n_{j} (m_j,T)$ the initial hadron gas particle density
and $\langle \sigma_{Dj} v_{Dj}\rangle$ is the particle specific average
flow and interaction cross-section.

\begin{figure}
\epsfig{width=7cm,clip=1,figure=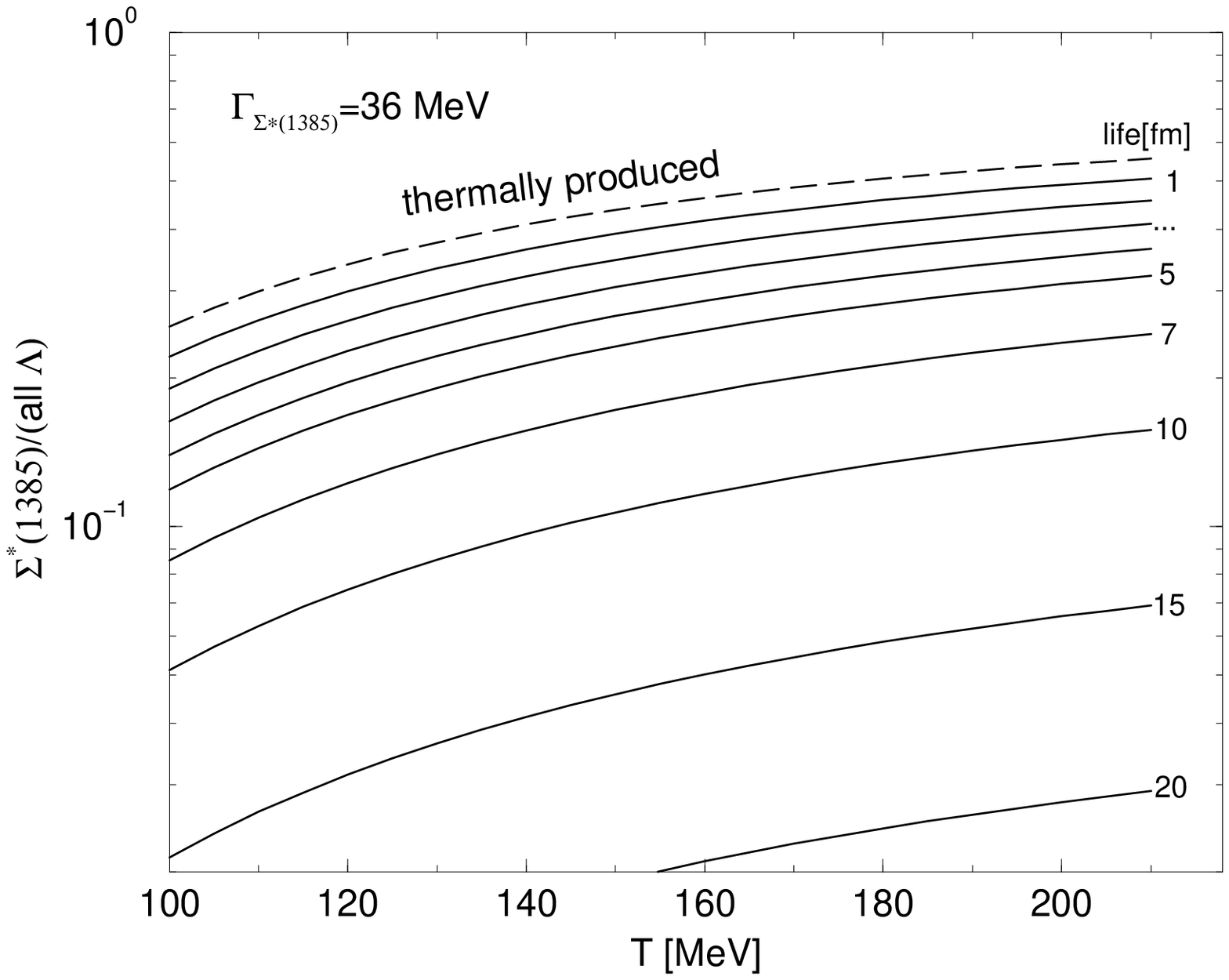}
\epsfig{width=7cm,clip=1,figure=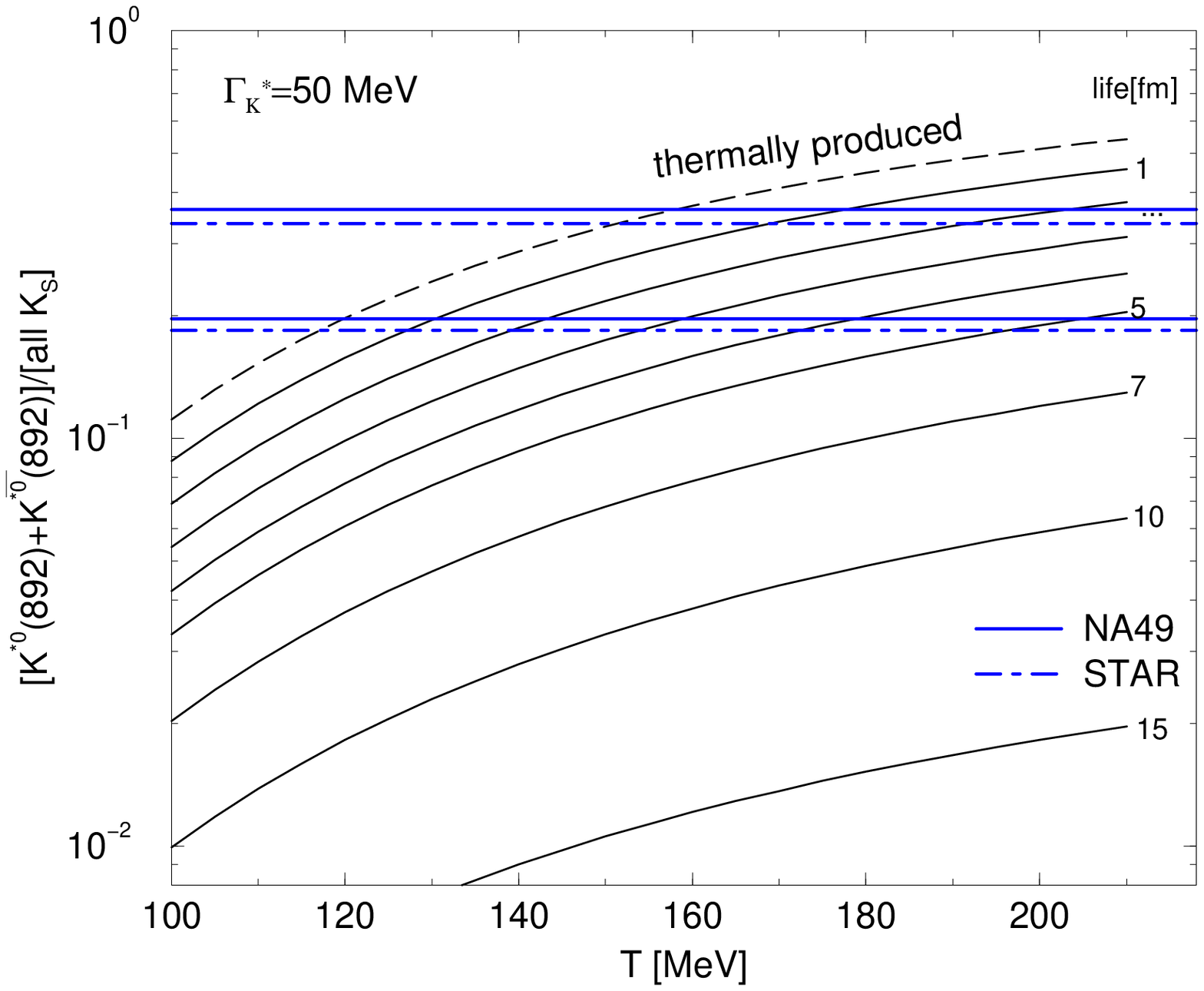}
\epsfig{width=7cm,clip=1,figure=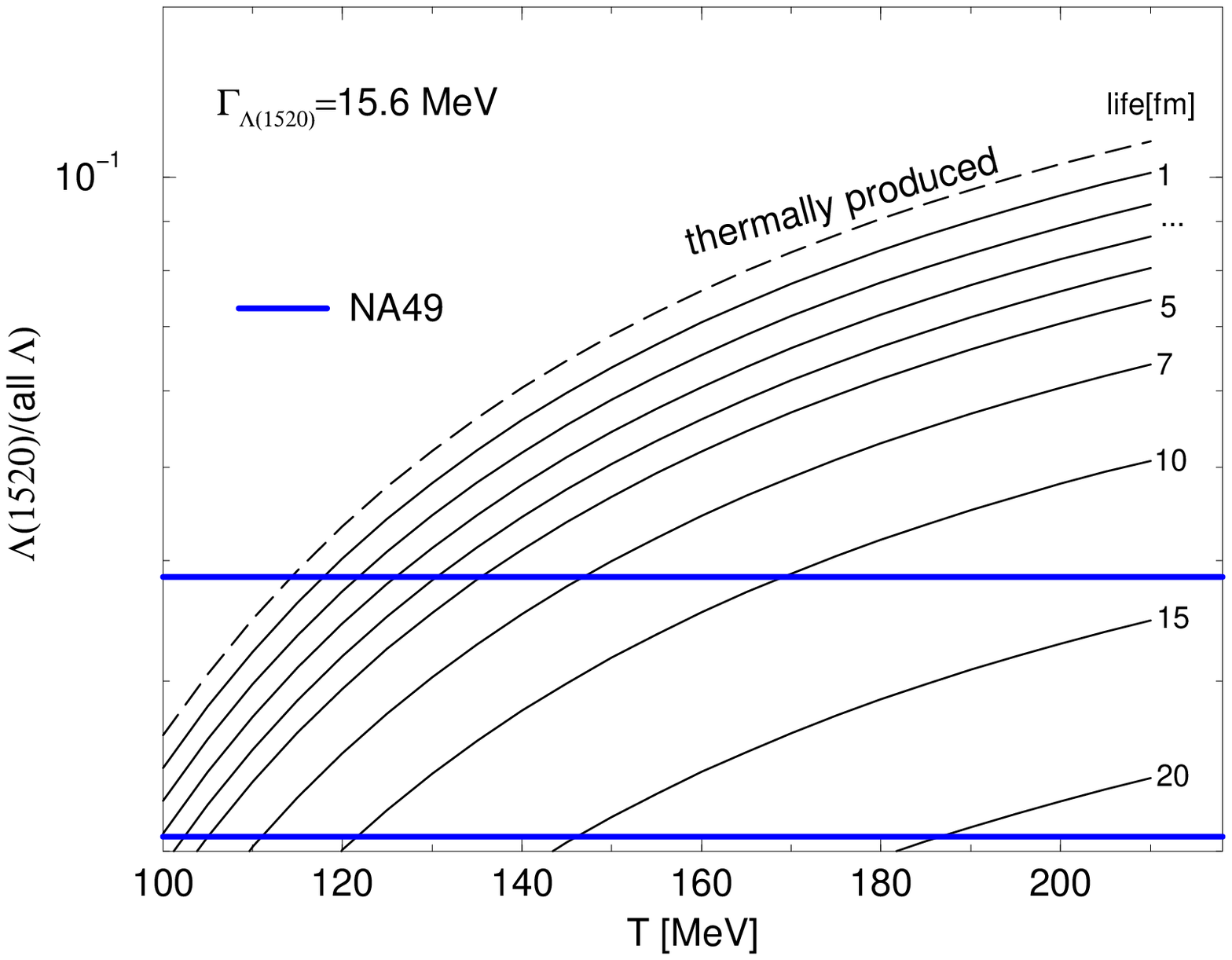}
\caption{ Observable relative resonance yields  as a 
function of temperature for a given 
length in time of  interacting hadron gas phase.
Horizontal and vertical lines give  experimental results.
\label{ratioT}}
\end{figure}

In this calculation, we have neglected the regeneration term 
$R\propto \langle \sigma_{Di}^{ INEL} v_{Di}\rangle  \rho_{i}$, 
since  detectable regenerated resonances
need to be real (close to mass-shell) particles.
Figure \ref{ratioT} shows how the ratios of 
$\Sigma^*/\Lambda$ and $K^*/K$ evolve
with varying hadronization time within this model.

\begin{figure}
\psfig{width=7.5cm,clip=1,figure= 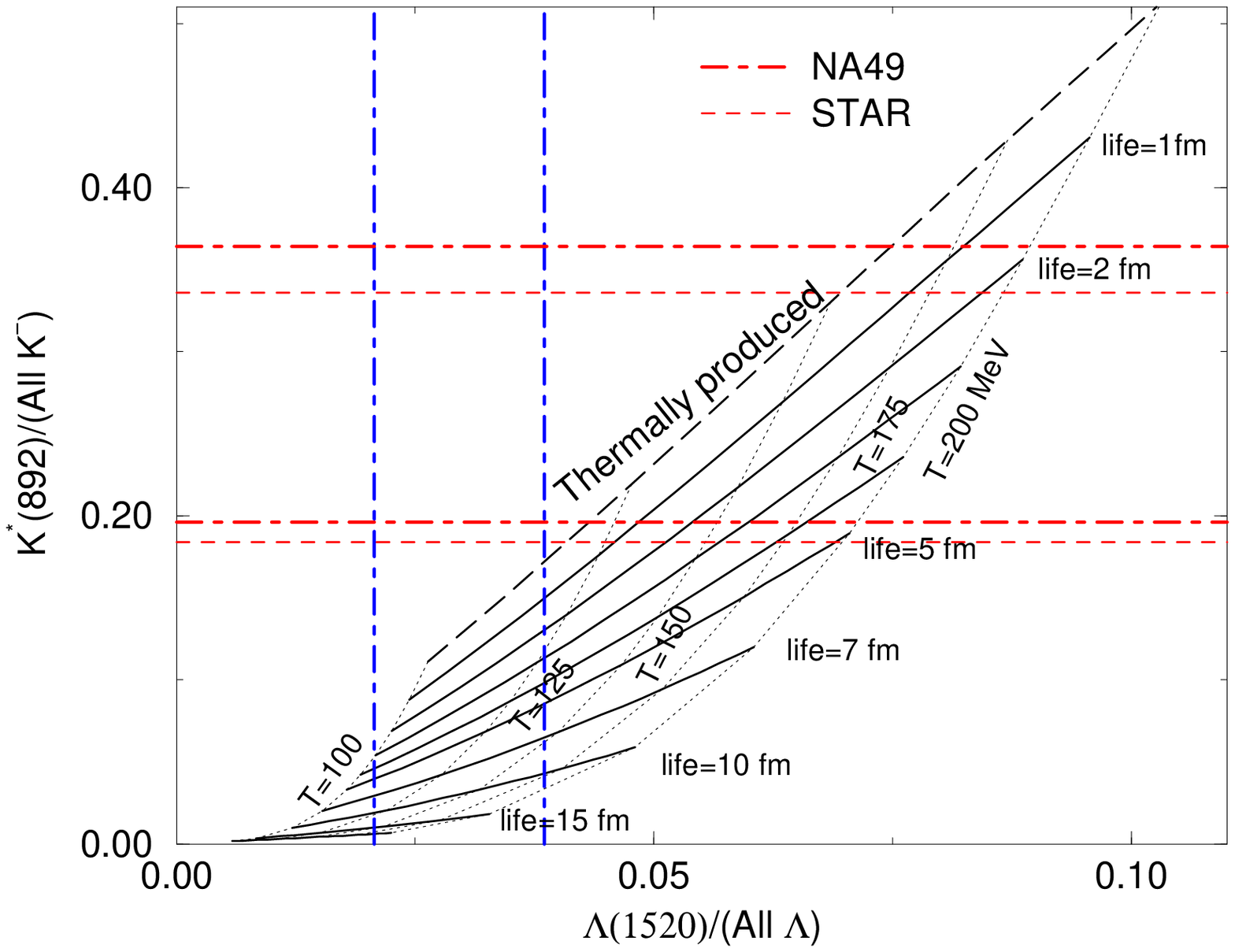}
\psfig{width=7.5cm,clip=1,figure= 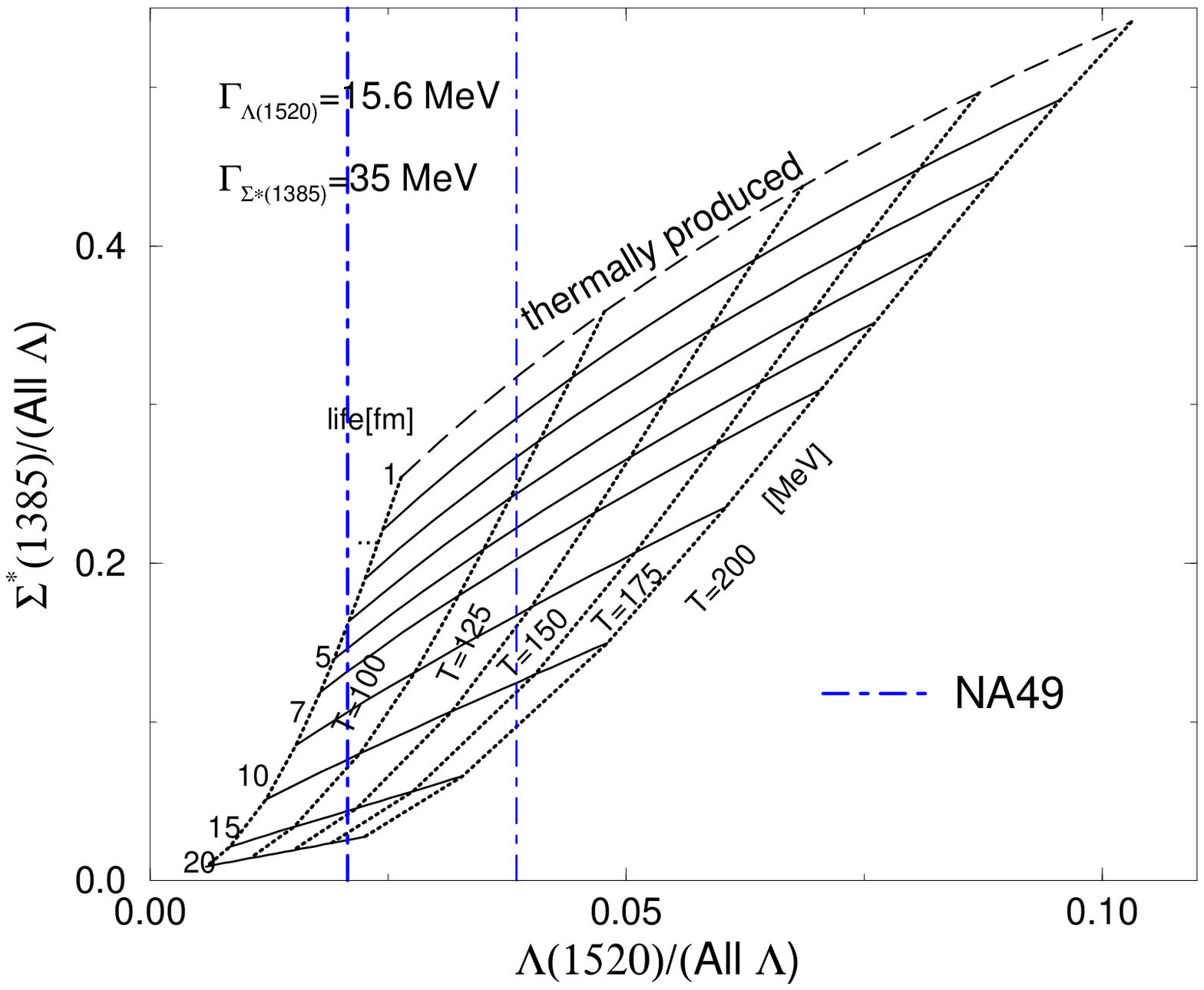 }
\psfig{width=7.5cm,clip=1,figure= 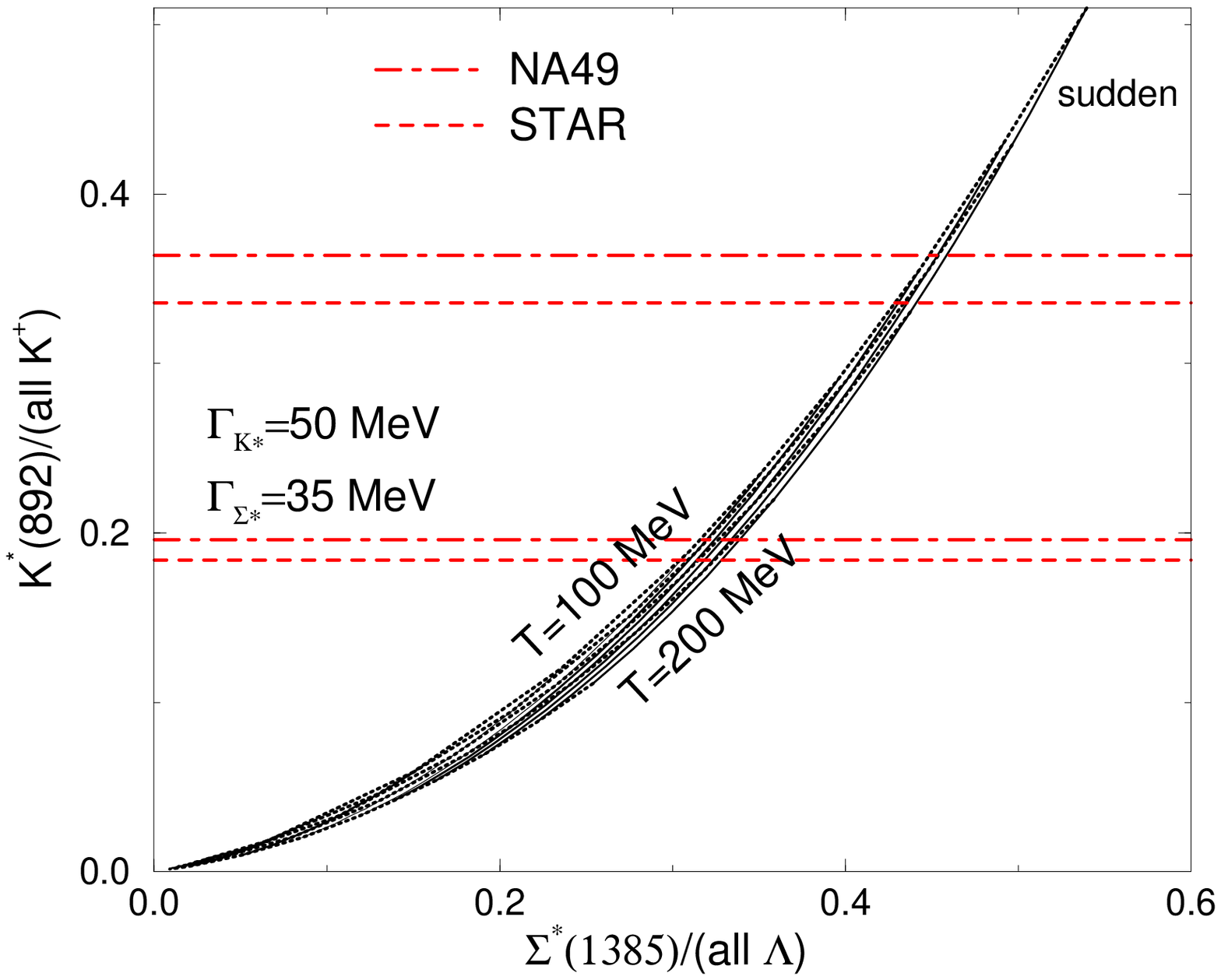 }
\caption{Projected combined relative yield diagrams over
a mesh of temperature and lifespan.
 Horizontal and vertical lines give  experimental results.
 \label{projdiag}}
\end{figure}

It therefore becomes apparent that measuring two such 
ratios simultaneously gives both the hadronization 
temperature and the time during which rescattering
is a significant effect.
Figure \ref{projdiag} shows the application of this 
method to $K^*,\Lambda(1520)$ and $\Sigma^*$.

The special role of  $\Lambda(1520)$ 
suppression is evident. $\Lambda(1520)$ is a
very peculiar resonance, since unlike the $K^*$ and $\Sigma^*$ it's extra
spin is believed to originate  from inter-quark orbital angular momentum (L=2).
It is therefore particularly susceptible to in-medium effects which suppress
it's yield or enhance it's width \cite{JRPRC,pasi}.
If this is the case, our model is able to account for existing observational
data and makes definite predictions for the measured $\Sigma^*$ abundance.

\section{Resonance spectra as direct  probe of freeze-out dynamics}

The near-independence of resonances on chemical potential makes 
their direct detection a powerful tool for examining further 
aspects of freeze-out dynamics.
It is apparent from the previous section that the ratio of the resonance
to the daughter particle with the same number of quarks is extremely
sensitive to freeze-out temperature.
\begin{figure}
\epsfig{width=7cm,height=7cm,clip=1,figure=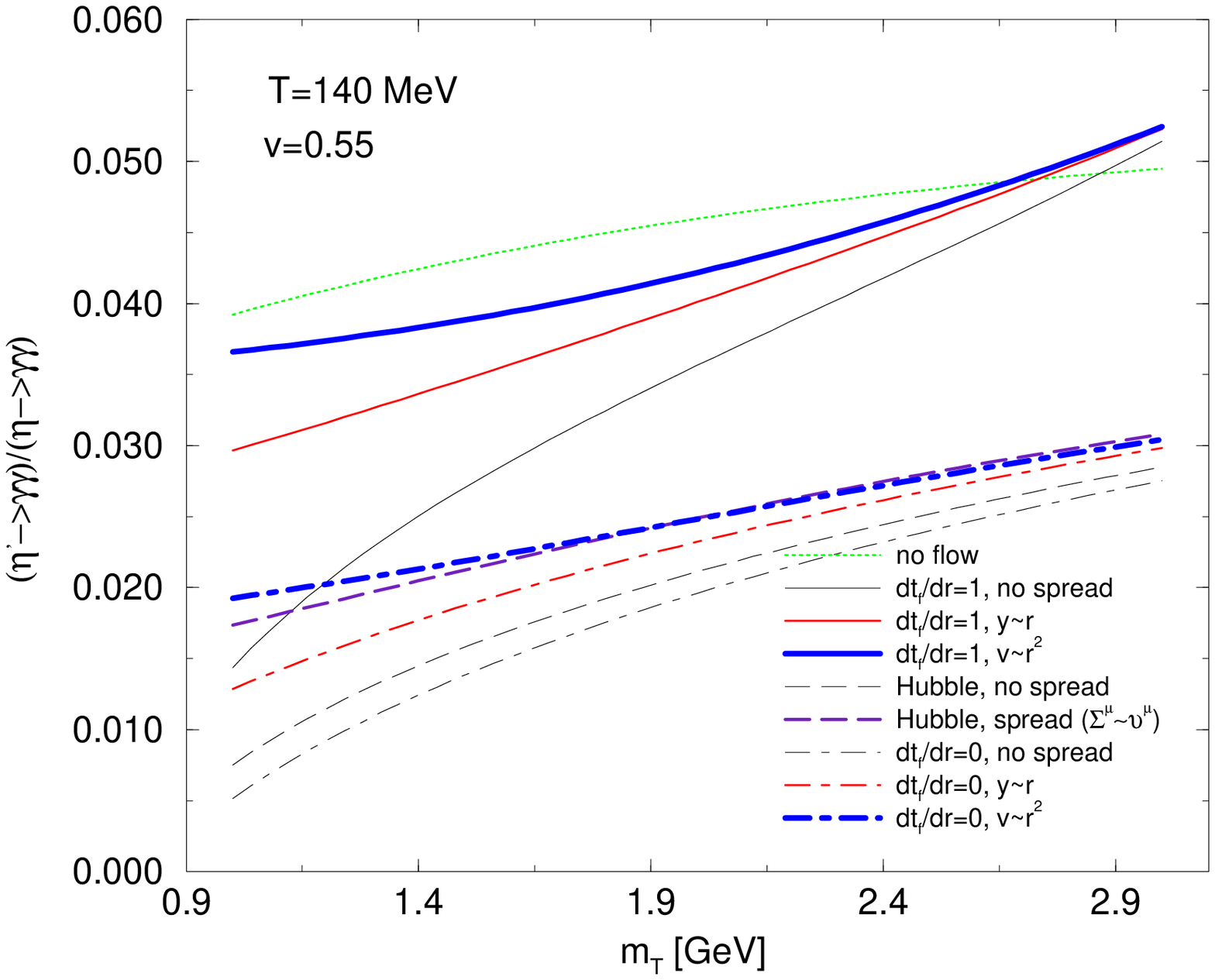}
\epsfig{width=7cm,height=7cm,clip=1,figure=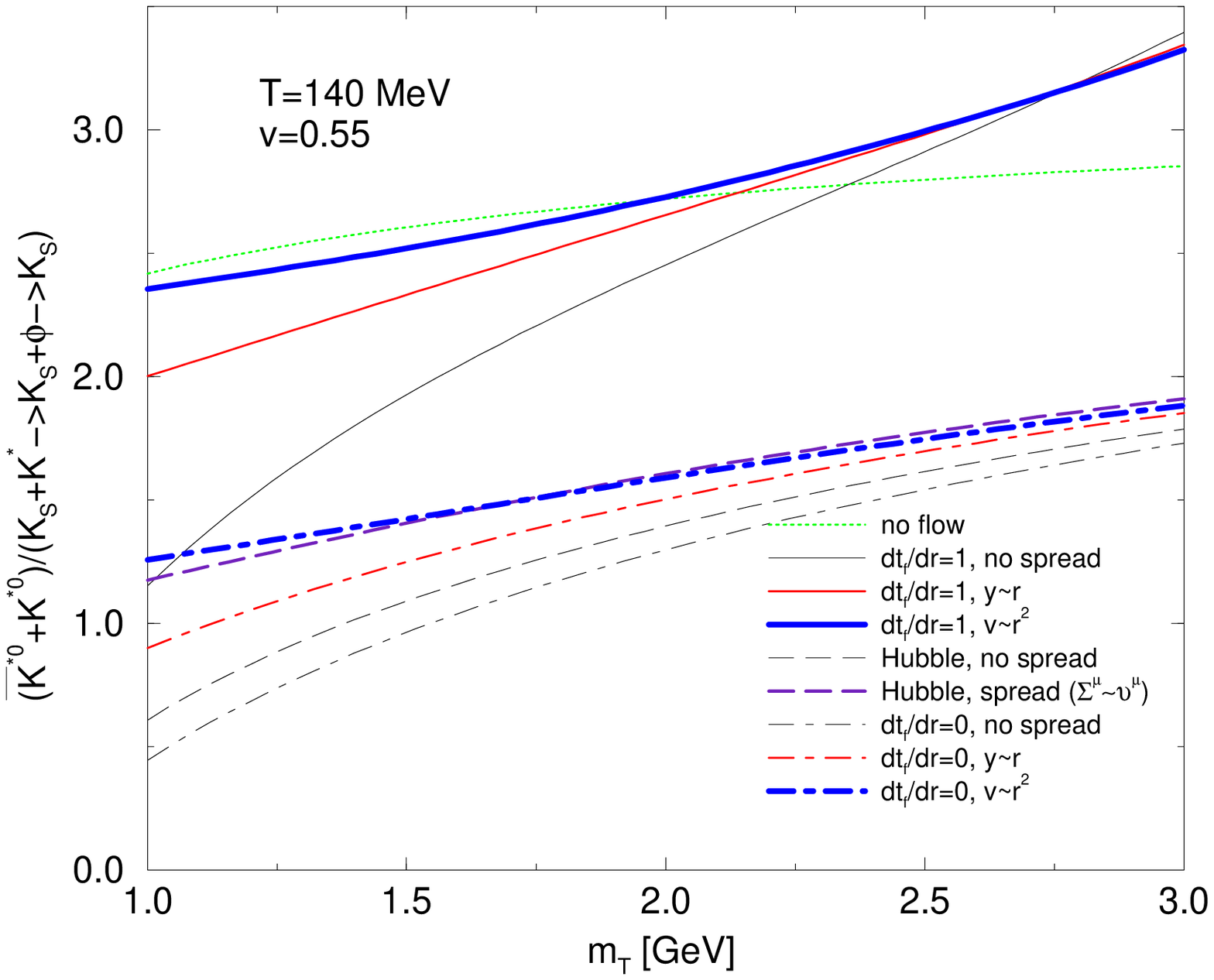}
\caption{Spectral ratios as function of $m_\bot$ for an 
array of
freeze-out surface conditions. \label{resratios}}
\end{figure}
If the $m_{\bot}$ dependence of this ratio
can be observed further freeze-out parameters can be
 extracted  \cite{statres}.   For a purely thermal source, in which 
\begin{equation}
\frac{dN}{dm_{\bot}} \propto g \cosh(y) m_{\bot}^2 e^{-\frac{m_{\bot}}{T}}
\end{equation}
the observed resonance ratio should be a step function, with $N^*/N=0$ for
$m_{\bot}<m^*$ and the ratio of degeneracies for the resonance
$g^*/g$ for  $m_{\bot}>m^*$.

However, a non-trivial flow profile and emission geometry can significantly
change this dependency.   For this reason, this ratio is an extremely sensitive
probe of both flow profile (the transverse flow as function of the radius)
and hadronization dynamics (how emission time varies with space, parametrized
by $\vec{v}_f$ in eq. \ref{cooperfrye}).
Figure \ref{resratios} shows this ratio (calculated by dividing two 
expressions of the form of eq. \ref{cooperfrye} but different masses) 
done for $\eta' \rightarrow \gamma \gamma/\eta \rightarrow \gamma \gamma$ 
and  $(K^*+\overline{K^*}) /K_S$.
It is apparent that this measurement is indeed a sensitive probe of freeze-out
dynamics and flow.   Moreover, flow effects are separated from
freeze-out dynamics, something which ``normal'' hadronic spectra can not
do effectively \cite{comparison}.

In conclusion, we have presented here an outline of the use of  resonances
 as diagnostic tools in the study of freeze-out dynamics.
We have shown that statistical fits to particle spectra require an admixture
of resonances consistent with thermal predictions, which strongly
suggests negligible post-hadronization rescattering dynamics.
We have described how the study of resonances yields and spectra 
can yield the hadronization 
temperature, the timescale of thermal freeze-out, 
and the freeze-out geometry  dynamics.
We expect these models to be constrained and developed further as resonance 
experimental data becomes more precise.

\acknowledgments
Supported  by a grant from the U.S. Department of
Energy,  DE-FG03-95ER40937\,. 



\begin{thebibliography}
\expandafter\ifx\csname natexlab\endcsname\relax\def\natexlab#1{#1}\fi
\expandafter\ifx\csname bibnamefont\endcsname\relax
  \def\bibnamefont#1{#1}\fi
\expandafter\ifx\csname bibfnamefont\endcsname\relax
  \def\bibfnamefont#1{#1}\fi
\expandafter\ifx\csname citenamefont\endcsname\relax
  \def\citenamefont#1{#1}\fi
\expandafter\ifx\csname url\endcsname\relax
  \def\url#1{\texttt{#1}}\fi
\expandafter\ifx\csname urlprefix\endcsname\relax\def\urlprefix{URL }\fi
\providecommand{\bibinfo}[2]{#2}
\providecommand{\eprint}[2][]{\url{#2}}

\bibitem{Fer50}
E. Fermi, {Prog. Theor. Phys.} {\bf 5}, 570 (1950).
\bibitem{Pom51}

I. Pomeranchuk, Doklady Akademii Nauk SSSR (in Russian)
(Proc. USSR Academy of Sciences)
{\bf 78}, 889 (1951).

\bibitem{Lan53}
L. D. Landau, {Izv. Akad. Nauk SSSR} (Ser. Fiz.) {\bf 17},
51 (1953); English edition, L. D. Landau,
{\it Collected Papers of L. D. Landau}, {\rm edited by D. Ter Haar},
Pergamon, Oxford (1965).

\bibitem{Hag65}
R. Hagedorn, {Suppl. Nuovo Cimento}  {\bf 3}, 147 (1965).

\bibitem{CooperFrye}
F. Cooper and G. Frye, {\it Phys. Rev.} {\bf D10} (1974) 186

\bibitem{jansbook}
J. Letessier and J. Rafelski, 
{\it Hadrons and Quark-Gluon Plasma}
(Cambridge University Press, Cambridge, 2002).

\bibitem{PBM99}
P. Braun-M\"unzinger, I. Heppe and J. Stachel, Phys. Lett. B
\textbf{465}, 15 (1999).

\bibitem{NA57}
L. Sandor, NA57 collaboration
J. Phys. G in press , see
       http://wa97.web.cern.ch/WA97/Publications.html
      {\it Hyperon production at the CERN SPS: results from the NA57
experiment}"

\bibitem{becattini}
F.~Becattini, M.~Gazdzicki and J.~Sollfrank,
Nucl.\ Phys.\ A {\bf 638}, 403 (1998);\\
F.~Becattini, J.~Cleymans, A.~Keranen, E.~Suhonen and K.~Redlich,
Phys.\ Rev.\ C {\bf 64}, 024901 (2001)
\\
F.~Becattini,
J.\ Phys.\ G {\bf 28}, 1553 (2002).

\bibitem{van-leeuwen}
S.V. Afanasiev et. al., NA49 Collaboration, Nucl. Phys. {\bf A715}, 161 (2003).

\bibitem{ourspspaper}
G.~Torrieri and J.~Rafelski,
J.\ Phys.\ G {\bf 28}, 1911 (2002)
[arXiv:hep-ph/0112195].

\bibitem{PBM01}
S.~V.~Akkelin, P.~Braun-Munzinger and Y.~M.~Sinyukov,
Nucl.\ Phys.\ A {\bf 710}, 439 (2002)

\bibitem{sqm2001}
G.~Torrieri and J.~Rafelski,
J.\ Phys.\ G {\bf 28}, 1911 (2002)

\bibitem{rafelski2002}
J.~Rafelski and J.~Letessier,
Nucl. Phys. A {\bf 715}, 98, (2003)\\
J.~Letessier and J.~Rafelski,
Int.\ J.\ Mod.\ Phys.\ E {\bf 9}, 107 (2000)

\bibitem{burward-hoy}
J. M. Burward-Hoy, PHENIX collaboration, 
Nucl. Phys. A {\bf A715}, 498, (2003).


\bibitem{castillo}
J. Adams et al., STAR Collaboration,
{\it Multi-Strange Baryon Production in Au-Au collisions at
$\sqrt{s_{NN}} = 130$ GeV}, nucl-ex/0307024, submitted to Phys. Rev. Lett.

\bibitem{PBMRHIC}
D.~Magestro,
J.\ Phys.\ G {\bf 28}, 1745 (2002)

\bibitem{bugaev_freeze}
K.~A.~Bugaev, M.~Gazdzicki and M.~I.~Gorenstein,
hep-ph/0211337.

\bibitem{florkowski}
W.~Broniowski and W.~Florkowski,\\
Phys.\ Rev.\ Lett.\  {\bf 87}, 272302 (2001)

\bibitem{fasthbt}
L.~P.~Csernai, M.~I.~Gorenstein, L.~L.~Jenkovszky, I.~Lovas and V.~K.~Magas,
Phys.\ Lett.\ B {\bf 551}, 121 (2003)
[arXiv:hep-ph/0210297].

\bibitem{sudden}
J.~Rafelski and J.~Letessier,
Phys.\ Rev.\ Lett.\  {\bf 85}, 4695 (2000)

\bibitem{heinzkolb}
U.~W.~Heinz and P.~F.~Kolb,
arXiv:hep-ph/0204061.

\bibitem{wrongHBT}
F.~Gastineau and J.~Aichelin,
arXiv:nucl-th/0007049.


\bibitem{fachini}
P.~Fachini,  STAR Collaboration,
Nucl. Phys. {\bf A715}, 462 (2003); J. Phys. G in press,  nucl-ex/0305034;
Haibin Zhang, STAR collaboration,
J. Phys. G in press,  nucl-ex/0305034;

\bibitem{vanburen}
G. Van Buren, STAR collaboration,
Nucl. Phys. A {\bf A715}, 129, (2003).

\bibitem{markert}
C.~Markert, STAR Collaboration,
J.\ Phys.\ G {\bf 28}, 1753 (2002).

\bibitem{friese}
V.~Friese, NA49 Collaboration,
Nucl.\ Phys.\ A {\bf 698} (2002) 487.

\bibitem{phasespace}
E. Byckling and K. Kajantie, {\it Particle Kinematics}, Wiley (1973).

\bibitem{mambo}
R. Kleiss, W.J.Stirling, Nucl Phys. B, \textbf{385} (1992) 413-432

\bibitem{resonances}
E. Schnedermann, J. Sollfrank and U. Heinz
NATO ASI series B,  Physics vol. \textbf{303} (1995)  175.

\bibitem{wa97}
F.~Antinori {\it et al.}  [WA97 Collaboration],
Eur.\ Phys.\ J.\ C {\bf 14}, 633 (2000).

\bibitem{shuryak}
E.~V.~Shuryak and G.~E.~Brown,
Nucl.\ Phys.\ A {\bf 717}, 322 (2003)
[arXiv:hep-ph/0211119].

\bibitem{JRPRC}
J.~Rafelski, J.~Letessier and  G.~Torrieri
Phys. Rev. C {\bf 64}, 054907 (2001), Erratum-ibid.C {\bf 65}, 069902 (2002).

\bibitem{pasi}
C.~Markert, G.~Torrieri and J.~Rafelski,
``Strange hadron resonances: Freeze-out probes in heavy-ion collisions,''
arXiv:hep-ph/0206260.

\bibitem{statres}
G.~Torrieri and J.~Rafelski,
Phys.\ Rev.\ C {\bf 68}, 034912 (2003)
[arXiv:nucl-th/0212091].

\bibitem{comparison}
G.~Torrieri and J.~Rafelski,
arXiv:nucl-th/0305071.

\end{thebibliography}
\end{document}